\documentclass[a4paper,nobibnotes,nofootinbib]{revtex4}

\usepackage{amsmath,amssymb}
\usepackage{epsfig}

\newcommand{\be}{\begin{equation}}
\newcommand{\ee}{\end{equation}}
\newcommand{\ba}{\begin{eqnarray}}
\newcommand{\ea}{\end{eqnarray}}
\newcommand{\bc}{\begin{center}}
\newcommand{\ec}{\end{center}}
\newcommand{\bfig}{\begin{figure}}
\newcommand{\efig}{\end{figure}}
\newcommand{\f}[2]{\frac{#1}{#2}}
\newcommand{\g}{\gamma}
\newcommand{\om}{\omega}

\newcommand{\no}{\nonumber \\}

\newcommand{\al}{\alpha}
\newcommand{\bal}{\bar \alpha}

\newcommand{\rr}[4]{#1, {\it #2 \/}{\bf #3} #4}

\begin{document}

\title{Confronting next-leading BFKL kernels  with  proton structure function 
data}

\author{R. Peschanski}
\email{pesch@spht.saclay.cea.fr}
\affiliation{Service de physique th{\'e}orique, CEA/Saclay,
  91191 Gif-sur-Yvette cedex, France\footnote{%
URA 2306, unit{\'e} de recherche associ{\'e}e au CNRS.}}
\author{C. Royon}\email{royon@hep.saclay.cea.fr} 
\affiliation{CEA/DSM/DAPNIA/SPP, F-91191 
Gif-sur-Yvette Cedex,
France}
\author{L. Schoeffel}\email{schoffel@hep.saclay.cea.fr} 
\affiliation{CEA/DSM/DAPNIA/SPP, F-91191 
Gif-sur-Yvette Cedex,
France}
\begin{abstract}
We propose a phenomenological study of the Balitsky-Fadin-Kuraev-Lipatov
(BFKL) approach applied to the data on  the proton structure 
function $F_2$ 
measured at HERA in the small-$x_{Bj}$ region. In a first part we use a 
simplified 
``effective kernel'' approximation  leading to   few-parameter 
fits   of $F_2.$ It  allows for   a  comparison between 
leading-logs (LO) and next-to-leading logs (NLO) BFKL approaches in the 
saddle-point approximation, using known resummed NLO-BFKL kernels. The NLO 
fits  
give  a  
 qualitatively satisfactory account of the running  
coupling constant effect  but quantitatively the  $\chi^2$ remains sizeably 
higher 
than  the 
LO fit at  fixed coupling. In a second part, a comparison of theory and data 
through a detailed 
analysis in Mellin space $x_{Bj} \to \om,$ leads to a more model independent
approach to the  resummed NLO-BFKL kernels we consider and  points out some 
necessary  improvements of the extrapolation at higher orders.
\end{abstract}

\maketitle

\section{Introduction}

Precise phenomenological tests of QCD evolution equations are one of the 
main 
goals of deep inelastic scattering phenomenology. For  the 
Dokshitzer-Gribov-Lipatov-Altarelli-Parisi
(DGLAP) evolution in  $Q^2$
\cite{dglap}, it has 
been possible 
to test it in various ways with  NLO (next-to-leading $\log Q^2$) and 
now
NNLO 
accuracy and 
it works quite well in a large range of $Q^2$ and $x_{Bj}.$
Testing precisely the Balitsky-Fadin-Kuraev-Lipatov
(BFKL) 
evolution in energy \cite{bfkl} (or  $x_{Bj}$) beyond leading order appears 
 more 
difficult. A  theoretically 
convenient way would be   to stay within the perturbative regime by using 
only 
massive 
or 
highly 
virtual colliding particles,  but   precision QCD phenomenology for the 
present day 
is provided mainly by the data \cite{data} on  deep-inelastic scattering at 
small 
$x_{Bj}.$

Indeed, the first experimental 
results from 
HERA 
confirmed the existence of a strong rise of the proton structure 
function 
$F_2$ 
with energy which, in the BFKL framework, can be   well described by a simple 
(3 
parameters) LO-BFKL fit
\cite{old,machado}. 
The main issue of Ref.\cite{old} was that not only the rise with energy but 
also 
the 
scaling violations  observed at small ${x_{Bj}}$ are encoded
in the BFKL framework through the $Q^2$ variation of the effective 
anomalous 
dimension. However one was led  \cite{old} to introduce an 
effective but unphysical  value of the  strong coupling constant  

$\al \sim .07-.09$ 
instead of 
$\al \sim .2$ in the $Q^2$-range considered  for  HERA small-${x_{Bj}}$ 
physics, 
revealing 
the need for NLO corrections. Indeed, the 
running of the strong coupling constant is not taken into account. 

 In fact,  the  theoretical task of computing these   
corrections 
appears to be quite hard. It is now in good progress but  
still under completion. For the BFKL kernel, they have been 
calculated after much efforts \cite{next}. In fact, they   
appeared to be so large 
that they 
 miss by a large amount the phenomelogical requirements and could even 
invalidate 
the whole theoretical
approach. 
Soon after, it was realized \cite{salam} 
that the main problem comes from the existence of spurious singularities 
brought 
together with the 
NLO corrections, which ought to be cancelled 
by an appropriate resummation at all orders of the perturbative 
expansion, resummation
required by consistency with  the QCD renormalization group. 

Indeed, various resummation 
schemes have been proposed \cite{salam,autres,lipatov} which satisfy the 
renormalization group 
requirements while retaining the computed  value of the 
NLO terms  in 
the 
BFKL kernel. Hence, the constraints can be  
satisfied and the next-to-leading 
order introduced without compromising the theoretical consistency of  
the BFKL scheme. 
However, 
the 
situation remains not so clear concerning   phenomenology\footnote{There 
exists fruitful 
phenomenological approaches starting from the DGLAP evolution and adding 
$\log 
x_{Bj}$ correction terms in the perturbative expansion \cite{thorne,altar}.}. 

To summarize the theoretical problems still 
remaining 
to be solved,  the determination of the impact 
factors associated 
with the coupling of the NLO-BFKL kernels with the virtual photon are 
still in progress \cite{bartels}. The factorization of the non perturbative 
coupling is also problematic at NLO-BFKL level. Another source of 
indeterminacy 
comes from  instabilities at 
 large energy of  the evolution equations which may 
lead 
to the dominance of  the   
non-perturbative ``Pomeron''
singularity \cite{instabilities}. 

On a phenomenological ground, we note that the 
resummation schemes  possess some ambiguity, since 
higher order logs (beyond the next-to-leading ones) are not known, apart 
from 
the 
renormalization group constraints. Such 
variations appear, {\it e.g.} in Ref. \cite{salam}, where  first
resummation schemes\footnote{ 
Other 
schemes have been 
proposed, e.g. 
\cite{lipatov}, which are left for further study.} have been proposed. In 
practice, 
we 
will  
consider the  schemes $S3$ and $S4$ of Ref. \cite{salam}
 together with  the  
one 
of Ref. \cite{autres}, in the formulation of  Ref. \cite{trianta},  including  
quark contributions to the kernel. 

The aim of our paper is  to compare the phenomenological efficiency of LO 
versus 
NLO BFKL approaches using  present precise experimental 
data on the  proton structure 
function 
$F_2$.  On 
the 
same footing it is also to  compare phenomenologically   the different NLO 
schemes  in 
order  
to check their  validity and  distinguish 
between different 
resummation options for the NLO-BFKL kernels.

For 
this sake, we shall use a two-step approach. First, we will consider a 
simpler 
version of the NLO-BFKL formulae by considering an ``effective kernel'' 
obtained 
using a well-known   ``consistency condition'' \cite{autres} satisfied by the 
NLO-BFKL kernels. Then, we formulate 
a 
saddle-point 
approximation for both 
cases  
allowing  a  similar formulation  for LO and NLO 
kernels\footnote{A 
more complete analysis will be possible within the same framework when the 
full NLO 
BFKL 
analysis including NLO impact factors
will be available 
.}. 

Our study is organized as follows. In Sec.\ref{2}, we 
present 
the construction of the effective kernels at LO and NLO levels and the 
corresponding  
saddle point  approximation for the 
proton 
structure function $F_2$.
In Sec.\ref{3}, we use these formulations to perform fits to the structure
 function data in the range ($x_{Bj} \le .01$ , $Q^2 \le 150\ GeV^2$)
suitable for a BFKL analysis. In Section \ref{4}
we perform a phenomenological determination of the Mellin transform   
$\tilde F_2 
(\om,Q^2)$ in the relevant range  and, with this information, we discuss the 
 ``consistency condition'' for the NLO-kernels 
under 
study. 
The final Section \ref{5} is devoted to a  summary, a discussion of our 
approximations  and an outlook. An   Appendix 
recalls 
the necessary formulae and
definitions  of 
the NLO kernels considered in this work.


\section{``Effective kernel'' and saddle-point approximation of BFKL 
amplitudes}
\label{2}

The BFKL formulation of the proton structure functions  can be formulated in 
terms of  
the double 
inverse Mellin 
integral
\begin{equation}
{F_2}=\int \! \int 
\frac {d\gamma d\om}{(2i\pi)^2} \left(\frac{Q^2}{Q_0^2}\right)^{\gamma} 
{x_{Bj}}^{-\om} \   {{\cal F}_2}(\g,\om)\ .
\label{bfkl0}
\end{equation}

At LO level one has (see, e.g. \cite{old}) 
\begin{equation}
{{\cal F}_2}(\g,\om)=\f {h_2(\g,\om)}{\om-\bal \ 
\chi_{L0}(\gamma)}
\label{bfkl}
\end{equation}
where $\bal \equiv  {\al_s N_c}/{\pi},$   $\al_s$ is the coupling 
constant which is merely  a parameter
at this LO level. The LO BFKL kernel is written as
\begin{equation}
\chi_{LO}(\gamma)=2\psi(1)-\psi(\gamma)-\psi(1-\gamma)\ .
\label{kernel}
\end{equation}
$h_2(\g,\om)$ is a prefactor  which  takes into account  both the  
phenomenological 
non-perturbative coupling to the proton and the   perturbative coupling 
to 
the virtual photon. Note that the  
variable 
$\gamma$  plays the role of a continuous 
anomalous dimension while $\om$ is the continuous index of the Mellin moment 
conjugated 
with the rapidity 
$Y\equiv \log {1}/{x_{Bj}}$. 

Recalling  well-known properties of LO-BFKL amplitudes, one  assumes that 
$h_2(\g,\om)$ is regular. The  pole contribution  at
$\om = \bal \ \chi_{L0}(\gamma)$
in  (\ref{bfkl}) leads  to a single Mellin transform in $\g$ for which one 
may use a 
saddle-point approximation 
at  small values of ${x_{Bj}}.$
Indeed, in the LO problem, it is known that 
the saddle-point approximation gives a very good account of the phenomenology 
as we will confirm later on. Going beyond the saddle-point approximation is  
theoretically more accurate, but leads to the same quality of fits, see, e.g. 
\cite{munierpesch}.
 
The saddle-point approximation 
gives
\begin{equation}
{F_2(x,Q^2)}\approx \ {\cal N}
\ 
\exp {\left\{\frac L2 + {\al_s} Y \chi_{L0}({\scriptstyle \frac 12}) -  
\frac 
{L^2}{2{\al_s} 
Y \chi''_{L0}(\frac 12)}\right\}}\ ,
\label{approxbfkl}
\end{equation}
where  $L\equiv 
\log({Q^2}/{Q_0^2})$ and ${\cal N}$ is a normalisation taking into account 
all 
the 
smooth prefactors\footnote{In particular, the square root prefactor   of 
the gaussian saddle-point 
approximation can be  merged in the normalization.}.  As a 
consequence, the 
only 
three relevant parameters in (\ref{approxbfkl}) are ${\cal N}, \bal$ and 
$Q_0.$ In this picture $\bal$ has to be considered as a
parameter and not a genuine  QCD coupling constant since the value obtained  
in 
the  fits  is not related to the coupling constant 
values in the 
considered  range of $Q^2.$  
As we shall now see, an effective 
saddle-point expression similar to (\ref{approxbfkl})
for 
the NLO BFKL analysis  can be written, but it will retain the  running 
property 
of the QCD coupling constant with its theoretically predetermined value at 
the 
relevant $Q^2$ range.
Hence it is no more a free parameter.

We will assume that, for high enough  virtuality $Q^2,$ NLO-BFKL solution 
for 
small-$x_{Bj}$ structure function is dominated by  the 
perturbative Green function. For this Green function  
\cite{salam,autres},  a 
consistency condition
relation holds, namely
 \be
 \omega-\f{\chi_{NLO}(\g,\omega)}{b\ L} = \omega-\al_{RG}(Q^2)\ 
{\chi_{NLO}( \g,\omega)}\equiv 0 \ ,
\label{al1}
\end{equation}
where 
\be
\left[\al_{RG}(Q^2)\right]^{-1} \equiv {b\ 
\log\left(Q^2/\Lambda_{QCD}^2\right)}\ ,
\label{al}
\ee
with $b=11/12-1/6\ N_f/N_c.$ 

The implicit relation (\ref{al1}) can be considered in two different 
ways. On the one hand, considering it as an implicit equation keeping $\om, 
Q^2$ 
fixed 
defining $
\g (\omega,\al_{RG}),$ it corresponds to 
 the saddle-point in the integration  of the Green function over $\g$. 
On the other hand, keeping $\g, Q^2$ fixed, it can be considered as an 
implicit 
equation 
for $\om 
(\g,\al_{RG}),$ and   appears as the NLO 
counterpart of the pole dominance from (\ref{bfkl}). 
Starting with the relation (\ref{al1}), one defines\footnote{The construction 
of 
the  
effective 
NLO BFKL kernel appears already in Ref.\cite{autres}.} an  effective NLO 
BFKL kernel
 \be
 \chi_{eff}(\g,\al_{RG})\equiv \f{ \om(\g,\al_{RG})}{\al_{RG}}\ .
\label{al2}
\ee
Using this kernel, and performing a saddle-point approximation at large $Y$ 
on the amplitude,  we 
obtain
\begin{equation}
{F_2(x,Q^2)}\approx \ {\cal N}
\ 
\exp {\left\{
\g_c \ L + {\al_{RG}}\ 
\chi_{eff}(\g_c,\al_{RG})\ Y  - 
\frac {L^2}{2\al_{RG}\chi''_{eff}(\g_c,\al_{RG})\ Y} \right\}}\ ,
\label{approxnlo}
\end{equation}
where $\g_c$ is defined by the implicit  saddle-point equation 
\be
\frac{\partial \chi_{eff}}{\partial \g}(\g_c,\al_{RG}(Q^2))= 0\ .
\label{gc2}
\ee
It is important at this stage to notice that the formula  
(\ref{approxnlo}) 
has only two free 
parameters 
${\cal N}$ and $Q_0$ instead of three for (\ref{approxbfkl}), once using 
the 
QCD universal 
coupling constant ${\al_{RG}}.$ It allows one to compare in a similar footing 
the LO 
and  NLO BFKL kernels to    $F_2$ 
data. In first place, it allows for   examining  the important 
effect 
of the running 
coupling constant on the fits.


\section{$F_2$ fits using LO and NLO BFKL kernels}
\label{3}

In practice, the determination of the  effective  kernel $\chi_{eff}(
\g,\al_{RG})$ using the implicit relation   (\ref{al2}) is made with the 
input corresponding to the 
(resummed)   NLO BFKL  kernels proposed  in the literature.
As recalled in the introduction, the NLO BFKL  kernels   have to be properly 
defined, 
in 
order to incorporate the next-leading terms calculated in \cite{next} 
and to 
get read of 
spurious singularities which would contradict the renormalization 
constraints 
\cite{salam,autres,lipatov}. There are different options for satisfying these 
constraints.
As an input of our analysis, we will concentrate on the schemes $S3$ and $S4$ 
of 
Ref. 
\cite{salam} and the $CCS$ scheme developped in Ref. \cite{autres}, following 
the 
version including quark contributions from Ref. \cite{trianta}.  The 
detailed formulae defining these kernels are explicitely given in the 
Appendix. We consider first the NLO schemes originally defined in 
Ref. \cite{salam} among which only the so-called 
$S3,S4$  
schemes are valid for phenomenological use. We also consider the resummation 
scheme of Ref. \cite{autres}, as developed in Ref. \cite{trianta} by 
including 
the quark contribution and denoted $CCS.$ The same 
procedure can be easily extended to other solutions for NLO BFKL kernels 
proposed in the 
literature. 
 
\begin{figure} [ht]
\begin{center}
\epsfig{figure=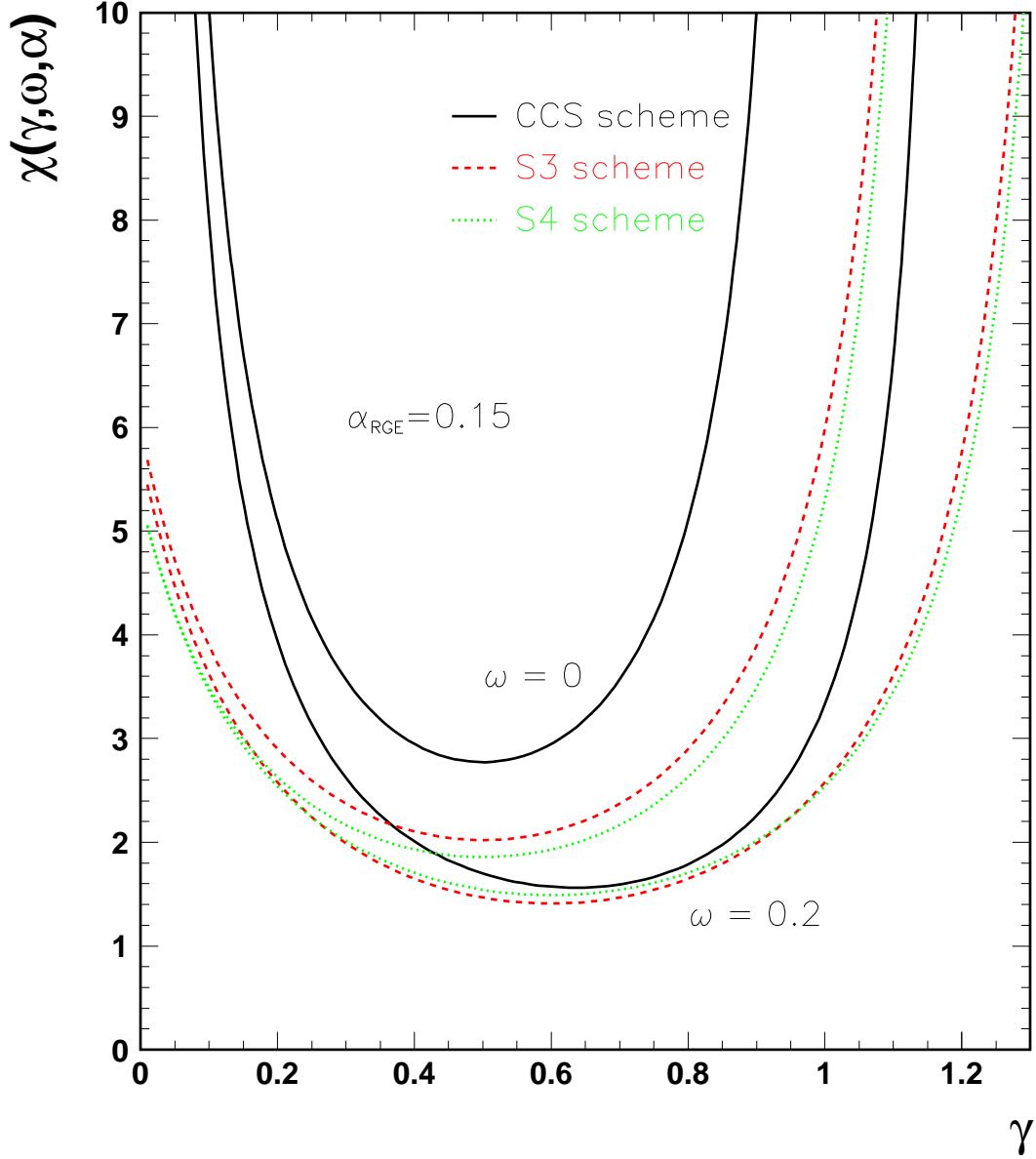,height=7.in}
\end{center}
\caption{{\it $\chi_{NLO}$ as a function of $\gamma.$} The curves correspond 
to 
$\alpha 
= 
cte=0.15$ for two values of $\om.$ Dark lines: $CCS$  scheme; dashed  lines: 
$S3$  
scheme; dotted lines: $S4$  scheme.}
\label{chinloalpha}
\end{figure}

\begin{figure} 
\begin{center}
\epsfig{figure=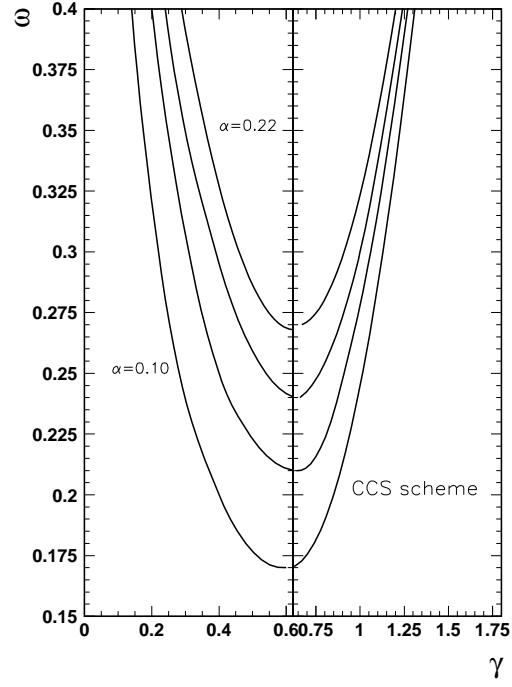,height=4.in}
\epsfig{figure=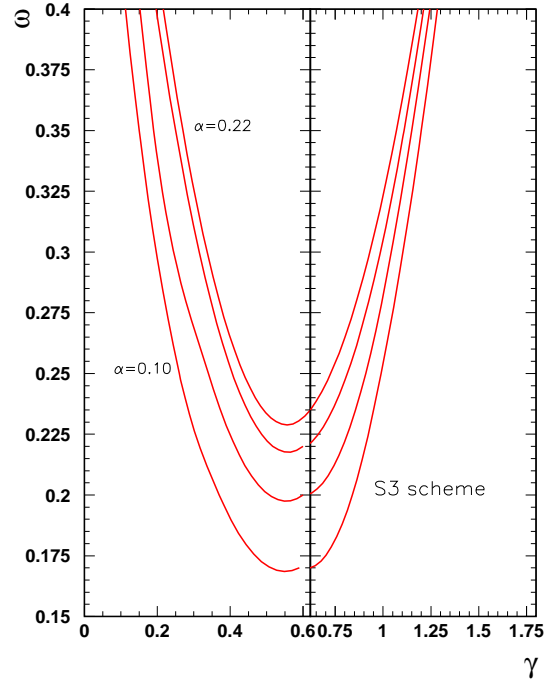,height=4.in}
\end{center}
\caption{{\it $\omega$ as a function of $\gamma$ for different values of
$\al_{RG}.$} Left:  $CCS$ scheme. Right:  $S3$ scheme, see text. 
$\alpha_{RG}$ 
varies 
between 0.1 to 0.24 by 
steps of 0.02.
For each value of $\omega$ in Eq.(\ref{al1}), two values of $\gamma$ are 
found and reported at 
each side of the minimum.}
\label{chinloalphab}
\end{figure}

In Fig.\ref{chinloalpha}, we show for the three 
different 
schemes 
$\chi_{NLO}(\g,\om)$ as a function of
$\gamma$ for $\al_{RG}=0.15$ and for different  values of 
$\omega$.
 In 
Fig.\ref{chinloalphab}, one finds the obtained 
kernels 
$\chi_{eff}(\g,\al_{RG})$ for the $CCS$ and $S3$ schemes\footnote{The 
$S4$ 
scheme gives an 
 effective kernel undistinguishable from  $S3.$}, after solution 
of 
the implicit 
equation  (\ref{al1}) for $\om(\g,\al_{RG}).$
 
\begin{figure} [ht]
\begin{center}
\epsfig{figure=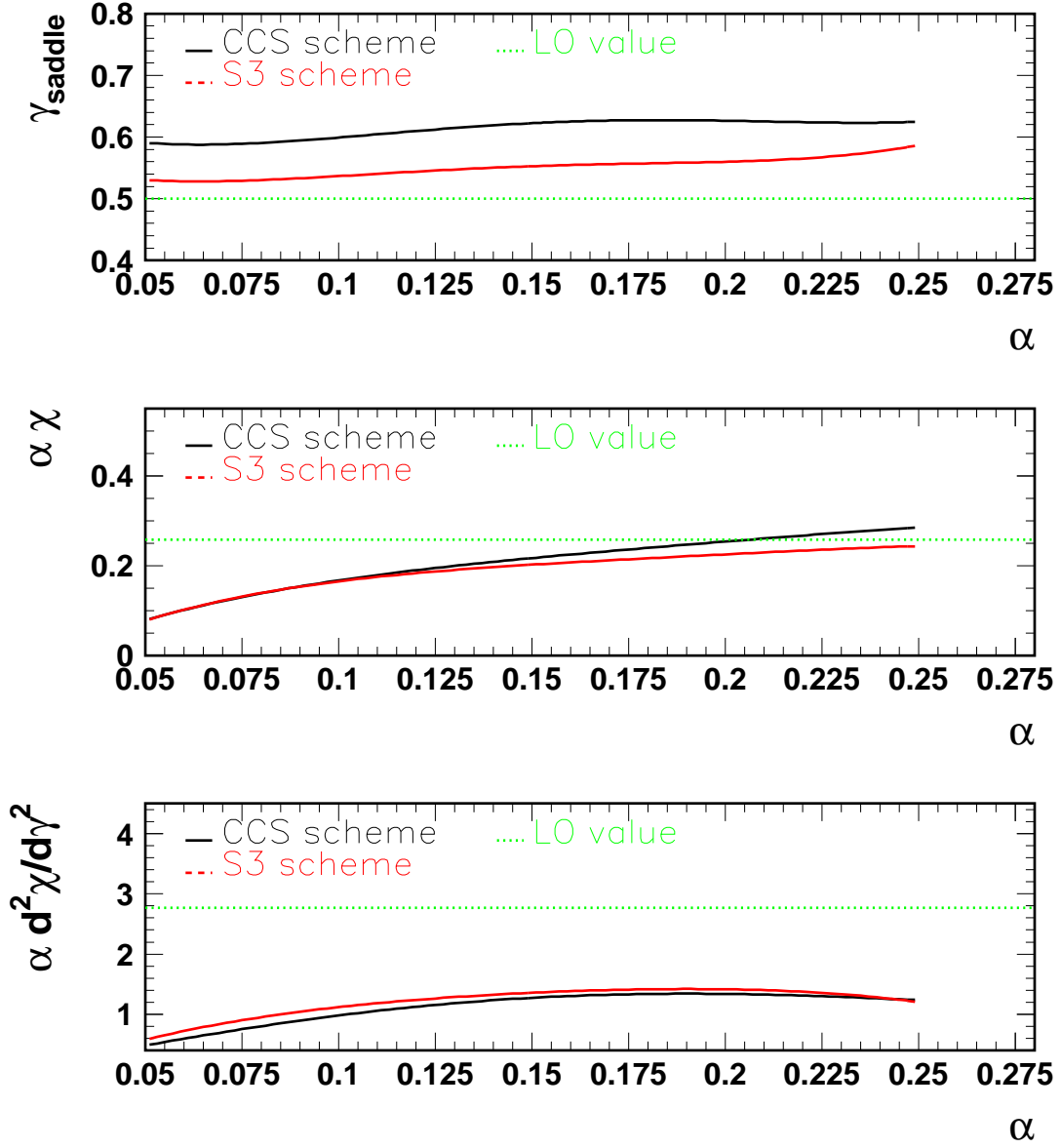,height=7.in}
\end{center}
\caption{{\it Determination of the saddle-point variables as functions 
of 
$\al_{RG}.$} Top: 
the saddle-point value $\g_c;$ Middle: the effective Pomeron intercept 
$\al_{RG}\ \chi_{eff};$ 
Bottom: the effective diffusion variable $\al_{RG}\ \chi''_{eff}.$ The LO 
fixed values 
correspond to the fit of section III.}
\label{fig_fitx_1}
\end{figure}

In Fig.\ref{fig_fitx_1}, one displays the parameter-independent values 
obtained for the set 
$\{\g_c,\al_{RG}\ \chi_{eff}(\g_c),\al_{RG}\ \chi''_{eff}(\g_c)
\}$ 
as a function of 
$\al_{RG}.$ They result from a numerical analysis of the NLO effective 
vertices (for the $CCS$ 
and $S3$ schemes) in the vicinity of the saddle-point. Knowing the 
dependence 
$\al_{RG}(Q^2),$ 
they allow one to predict the behaviour of $F_2,$ up to the determination of 
the 
free parameters 
$\cal N$ and $Q_0,$ see (\ref{approxnlo}). By comparison we show the 
corresponding values taken 
by the parameter-dependent set $\left\{\g_c\equiv {\scriptstyle \f 
12},\bal\chi_{LO}(\g_c),\bal\chi''_{LO}(\g_c)\right\}$ which are the 
ingredients of the LO 
formula 
(\ref{approxbfkl}). 
It is 
interesting to note that the hard pomeron intercept  
$\al_{RG}\ \chi_{eff}(\g_c)$ is compatible 
with the LO fitted value in a physical range of $\al_{RG} \sim .2$ 
while 
the LO value $\bal 
\sim .09$ is not physically motivated.  

Let us come now to the 
quantitative analysis.

The parameters  of the fits are given in Table I. 
Note that for the LO quantities 
$\bal$ is a fitted constant  whereas in the NLO ones it is given by the standard 
renormalization group formula (\ref{al}). However for completeness and inspired by  
theoretical arguments presented in Refs. \cite{thorne,altar,instabilities}, we have also 
considered the LO expression (\ref{approxbfkl}), with the constant $\al_s$ replaced by 
the running coupling  (\ref{al}). The corresponding fit parameters are also given in 
Table I (referred as LO'(${\alpha}_{RG}$)).

\begin{figure} 
\begin{center}
\epsfig{figure=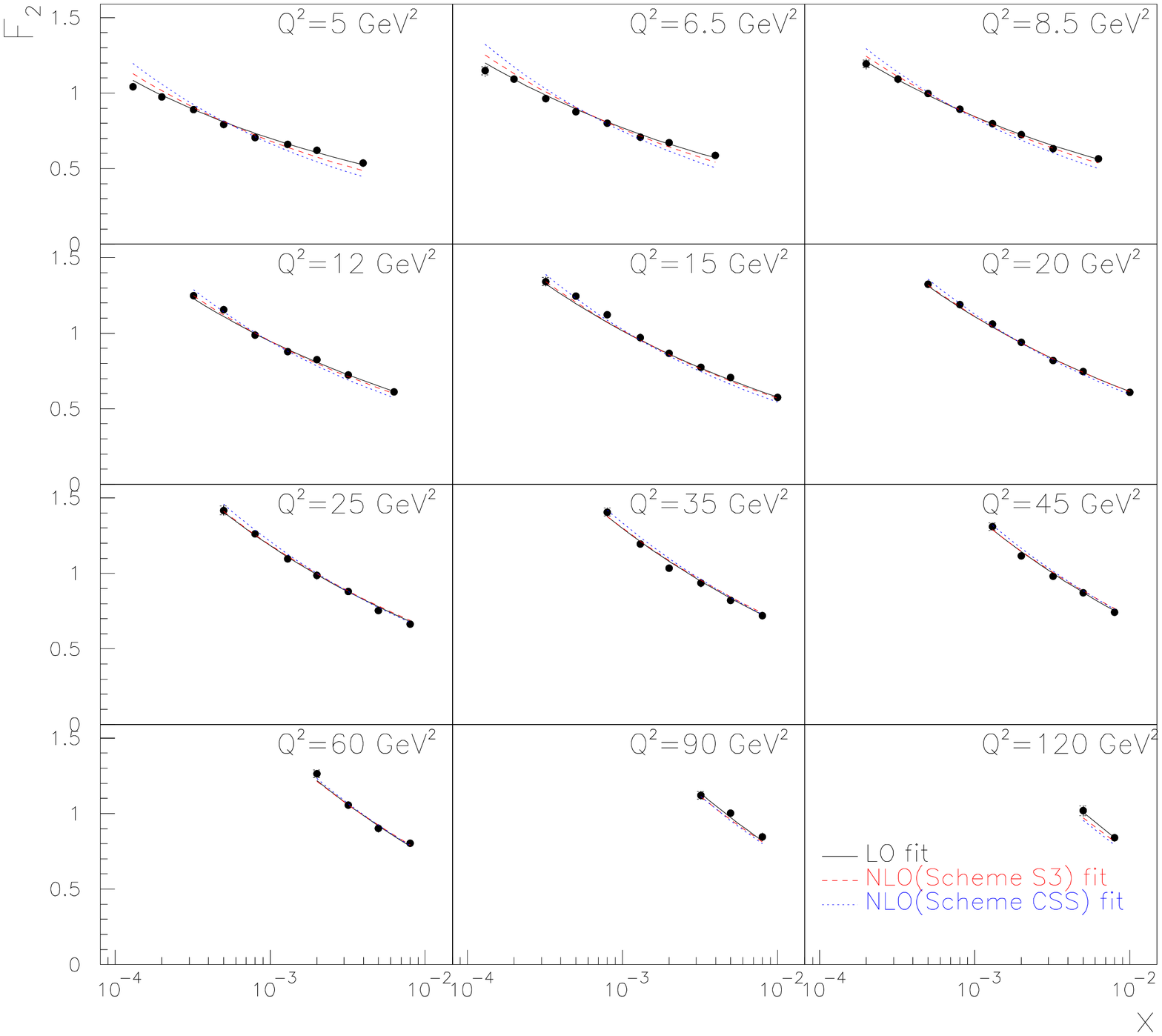,height=7.in}
\end{center}
\caption{{\it Results of the BFKL fits to the H1 data.}
LO BFKL kernel (continuous lines); NLO BFKL kernels ($S3$: Dashed lines, 
$CCS$: 
Dotted 
lines).
 }
\label{fig_fitx_2}
\end{figure}

In Fig.\ref{fig_fitx_2}, we display the results of the BFKL fits to H1 data 
for  LO and the $S3$ and $CCS$  schemes at NLO. As is clear from the figure 
the LO 
fit 
is doing a much better job than the considered NLO schemes (even considering 
 that the NLO kernels depend on one less parameter). While the 
qualitative behaviour is correct  they fail 
to take into account the quite precise $F_2$ data, especially at lower $Q^2$ 
where they show a too steep behaviour in $x_{Bj}.$ Note that the $S3$ scheme 
is somewhat better as confirmed by the value of the $\chi^2,$ displayed in 
Table 
I.
As obvious from Table I, the LO'(${\alpha}_{RG}$) fit (not represented in the 
figures) is not successful either.

\begin{figure} 
\begin{center}
\epsfig{figure=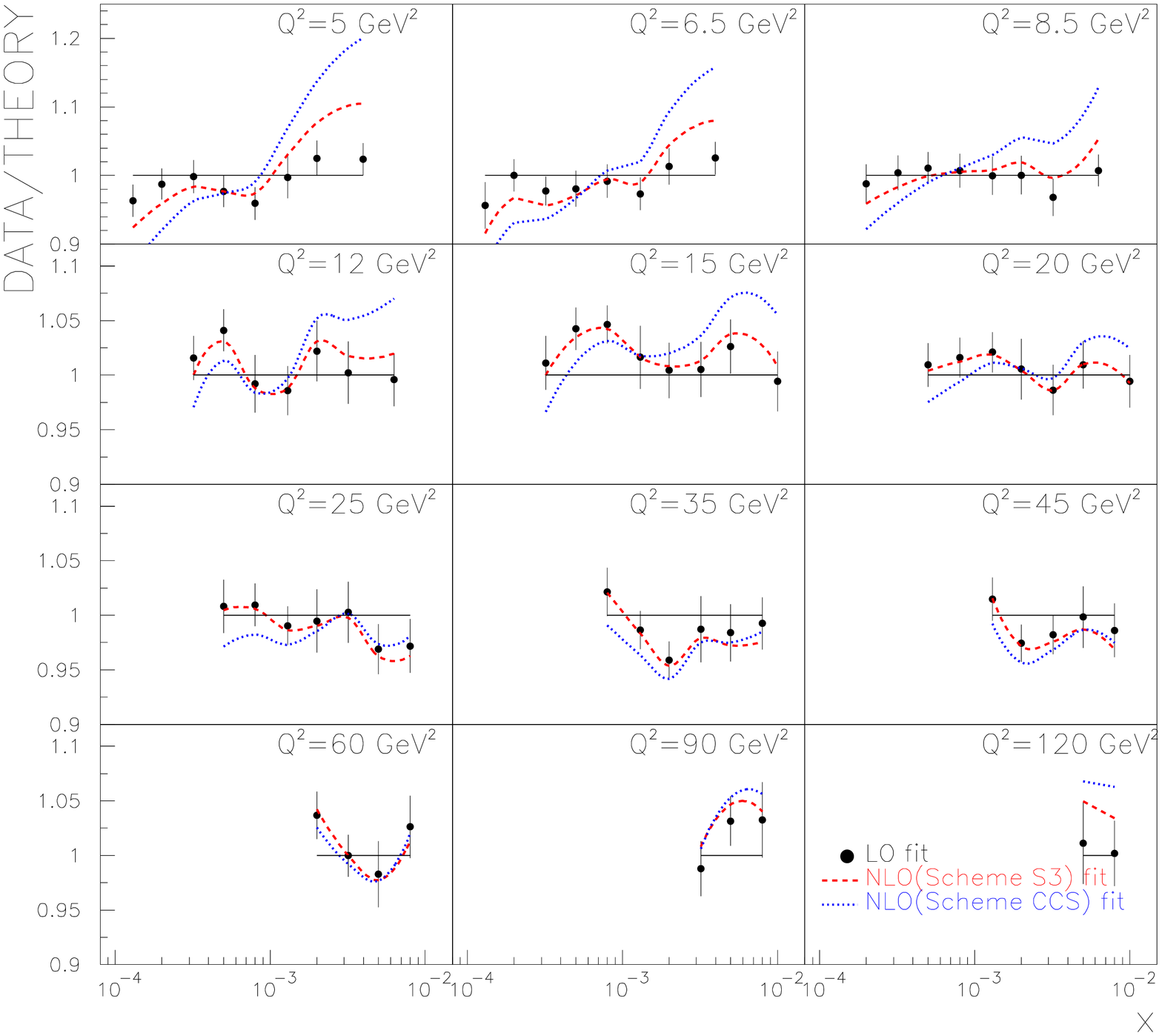,height=7.in}
\end{center}
\caption{{\it Ratio theory/data for the  BFKL fits to the H1 data.}
LO BFKL:  points with error bars; NLO BFKL Scheme $S3$: Dashed line; NLO BFKL 
Scheme $CCS$: Dotted line.
 }
\label{ratio}
\end{figure}

In order to show the behaviour of the different fits with more detail, we 
display
in Fig.\ref{ratio} the comparison of the ratio theory/data for the LO fit 
(with the expected error bars) and the NLO ones. This figure clearly confirms 
the problems at lower $Q^2.$

\section{Analysis in Mellin space}
\label{4}

In this section we want to analyze in more detail the features of the 
BFKL 
parametrizations and in particular the reasons of the still quantitatively 
unsatisfactory 
results of the NLO fits. For this sake, it is important to come back to 
the 
key ingredient of our analysis, i.e. the dominance of the hard Pomeron 
singularity expressed by the relation (\ref{al1}).  As mentionned in 
section 
II, this relation is expressed in Mellin space, and our aim is now to 
make a 
phenomenological test of this relation directly in the $\om$ space, 
without 
using the approximation of  effective kernels.

Equality  (\ref{al1})  can be checked at NLO using the GRV98 
\cite{GRV98}, MRS2001 \cite{MRS2001}, CTEQ6.1 \cite{CTEQ6.1} and ALLM 
\cite{ALLM}
parametrisations. These four parametrisations give a fair description of 
the 
proton
structure functions measured by the H1 and ZEUS collaborations over a 
wide 
range of $x_{Bj}$ and $Q^2$, as well as fixed target experiment data. The 
three first
parametrisations correspond to a DGLAP NLO evolution whereas the ALLM 
one
corresponds to a Regge analysis of proton structure function 
data\footnote{We introduced the ALLM parametrisation in order to avoid 
biases which could be due to the DGLAP constraints. In fact no 
significant 
difference appears in the resulting analysis, at least at moderate $Q^2$ 
where our 
analysis is being done.}. Since these
parametrisations give a good description of data, we can use them to 
test
easily the properties of the NLO BFKL consistency condition  (\ref{al1}). 
This allows us to make a direct computation of the Mellin
transform $\tilde F_2(\om,Q^2)$ of the proton structure function and 
study the 
NLO BFKL properties in
Mellin space.
Writing the Mellin transform as (cf. formula (\ref{bfkl0}))
\begin{eqnarray}
\tilde F_2(\om,Q^2)\equiv \int 
\frac{d\gamma}{2i\pi} \ \left(\frac{Q^2}{Q_0^2}\right)^{\gamma} 
\  {{\cal F}_2}(\g,\om) \ ,
\label{tilde}
\end{eqnarray}
it is easy to realize that
\begin{eqnarray}
\gamma^*(\om,Q^2) = \frac{{ d} \ln \tilde F_2(\om, Q^2)}{{ d} \ln Q^2}\ ,
\label{gamma}
\end{eqnarray}
where $\g^*$ is the saddle-point for the structure function, or in other 
well-known terms its {\it effective anomalous dimension} . Note that  
under our 
assumption that the saddle-point  of the gluon Green function is 
transmitted to 
the full amplitude (see previous section) $\g^* \approx \g_c$, and thus 
the 
relation (\ref{gamma}) allows for  a phenomenological approach of the 
consistency 
condition.

\begin{figure} [ht]
\begin{center}
\epsfig{figure=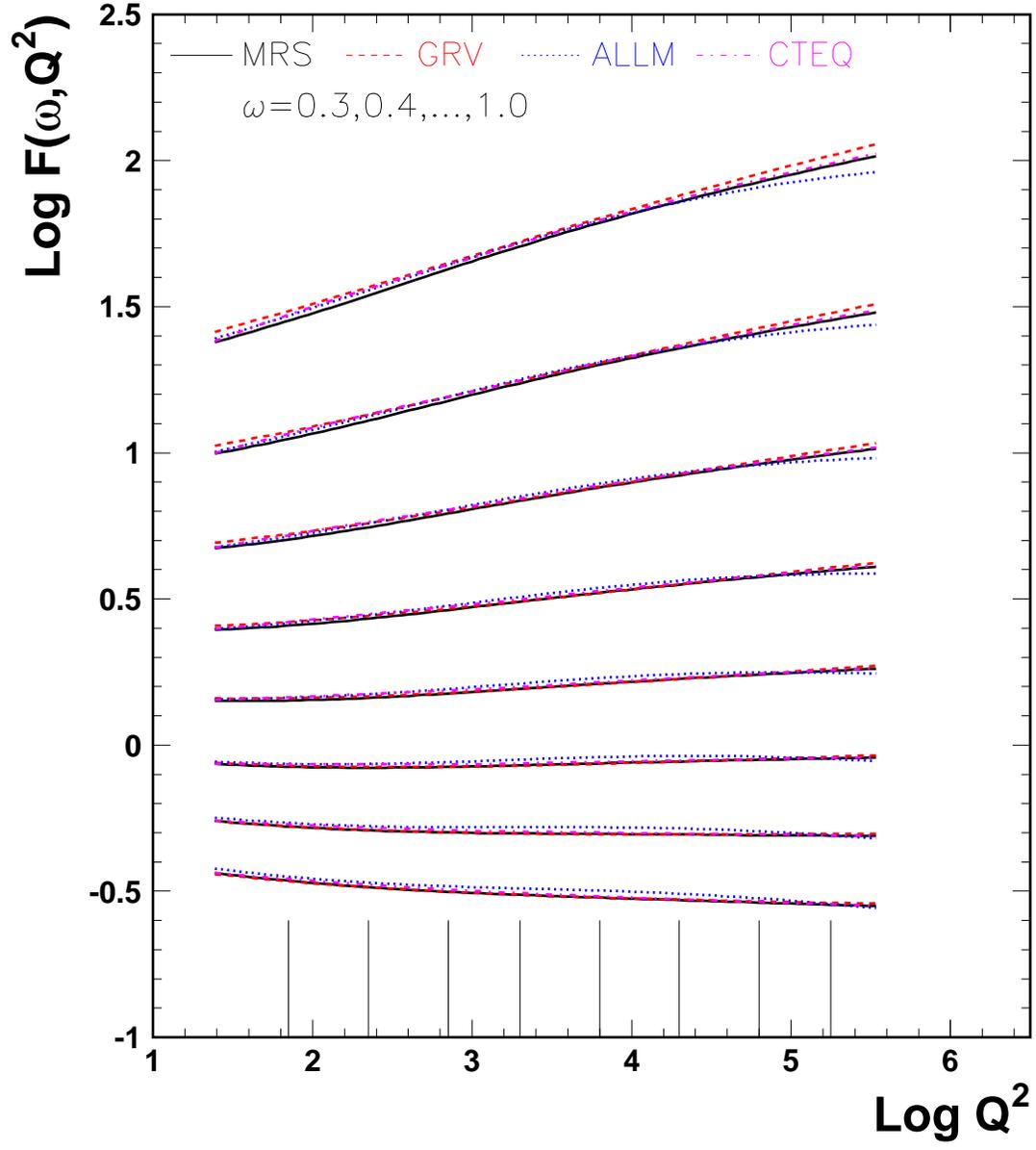,height=7.in}
\end{center}
\caption{{\it $\log F_2(\om,Q^2)$ as a function of $\log 
Q^2.$} 
Four different
parametrisations of the proton structure function have been considered: 
MRS2001 (continuous), GRV98 (dashed),  ALLM (dotted), CTEQ6.1 
(dotted-dashed).}
\label{logf2}
\end{figure}

In Fig. \ref{logf2}, we display $\log \tilde F_2(\om,Q^2)$ as a function of 
$\log Q^2$ 
for
the four parametrisations described above for different values of 
$\omega$.
The upper curve corresponds to $\om=0.3$, and $\om$ varies in steps of 
0.1 up
to the down curve which corresponds to $\om=1.0$. The limiting 
values\footnote{By 
various numerical tests, we checked that  limiting the Mellin transform 
range to 
$0.3<\om<1$  corresponds to 
take  into account  data points with
$ 10^{-4}<x_{Bj} < 10^{-2}$.} in 
$\om$ 
are due
to the fact that, one the one hand, we do not want to use data at too 
high $x_{Bj}$ to 
test BFKL
NLO properties and on the other hand, the data are lacking at very small 
$x_{Bj}$. We do not see large differences between
the  parametrisations used in this analysis. The slope in Fig.
\ref{logf2} gives directly the value of $\gamma^*$ according to formula
(\ref{gamma}). The vertical lines define the bins in $Q^2$ where 
the slope of $\tilde F_2(\om,Q^2)$ can be safely evaluated from the curves.

\begin{figure} [ht]
\begin{center}
\epsfig{figure=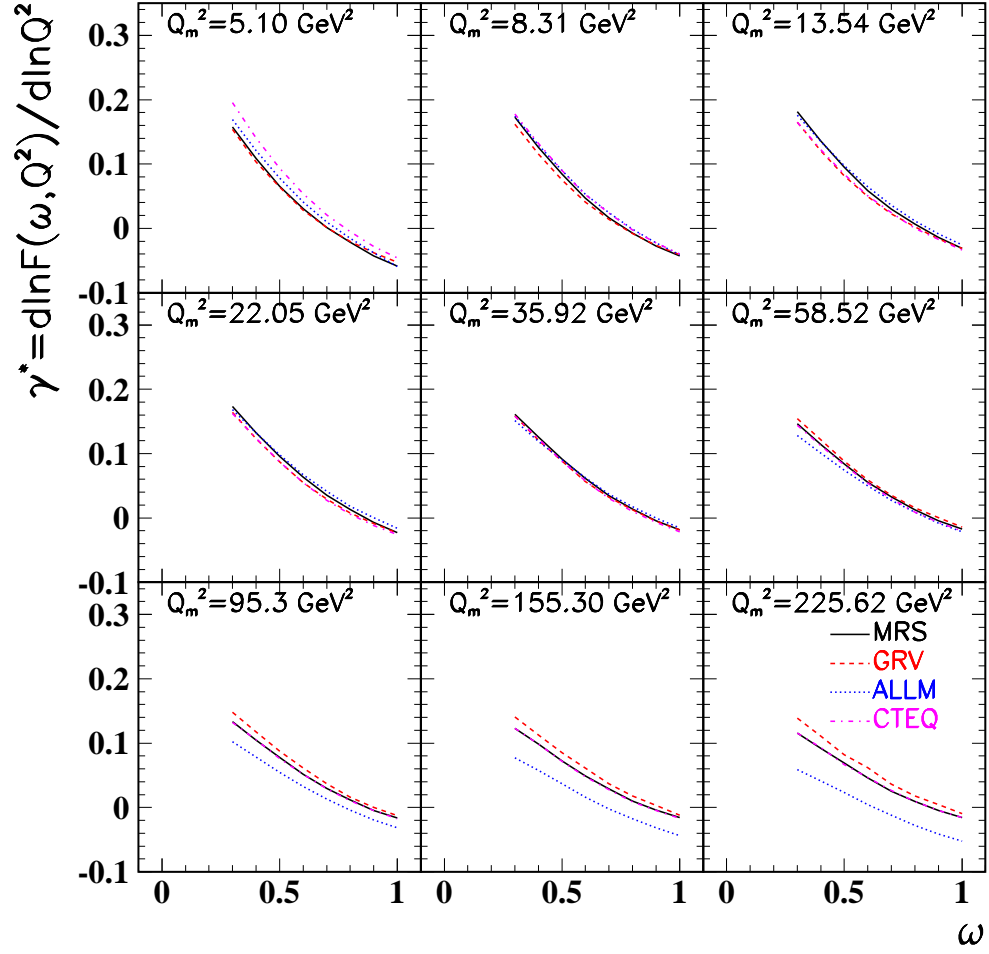,height=7.in}
\end{center}
\caption{Derivation of the anomalous dimension $\gamma^*(\om,Q^2)$
by computing the slope of $log F_2$ as a function of $\log Q^2$.}
\label{gammab}
\end{figure}

In Fig.\ref{gammab}, we derive  the values of $\gamma^*$ which are the 
slopes of 
$\log \tilde F_2$ in $\log Q^2$ in the different bins of $Q^2$ defined
above, for the four parametrisations. We do not see sizable differences 
between
the parametrisations except at higher values of $Q^2,$ which are anyway 
not used 
in our analysis. Hence, in the kinematical range appropriate for BFKL 
phenomenology, the different sets of structure functions, being or not 
driven by 
the DGLAP equations, do not give rise to noticeable differences.

\begin{figure}
\begin{center}
\epsfig{figure=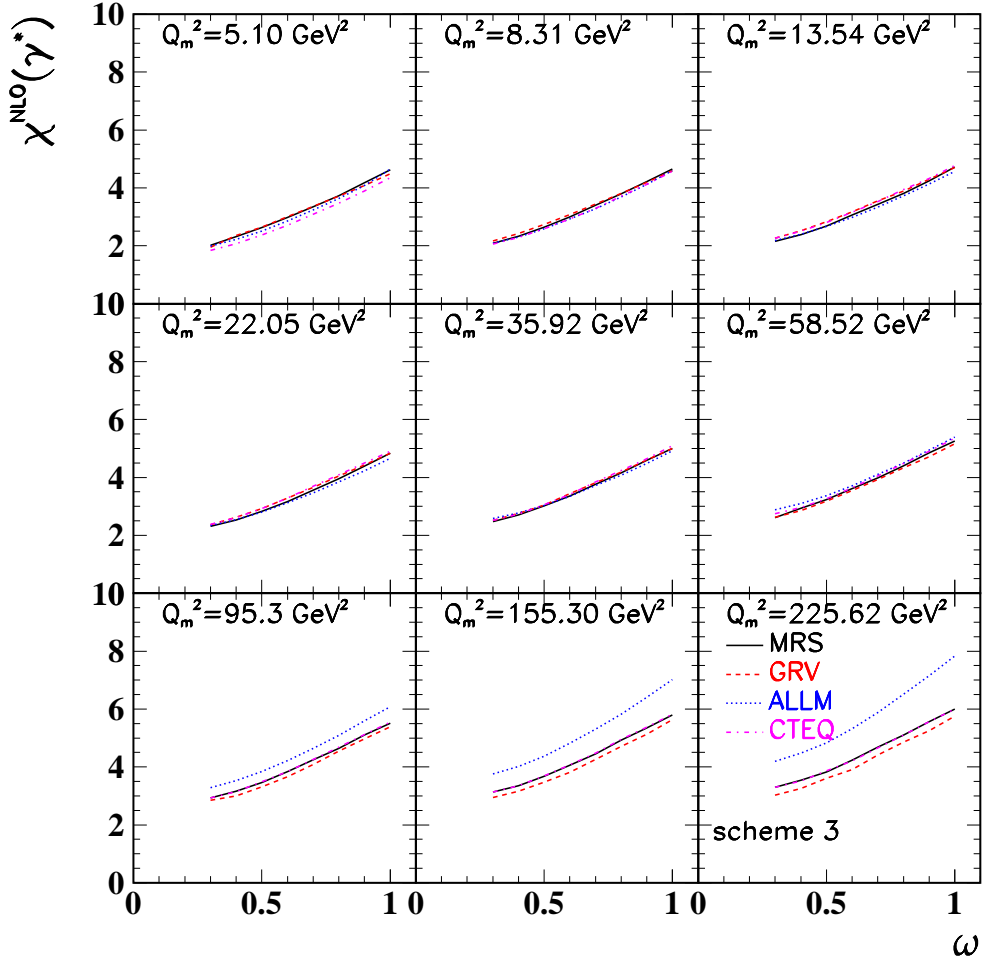,height=7.in}
\end{center}
\caption{{\it Test of $\chi(\om,Q^2) $ for scheme $S3.$} The 
results for the  
four parametrizations are shown for the different bins in $Q^2,$ (see 
text).}
\label{chinlos3}
\end{figure}

After having determined the values of $\gamma^*$, it is possible to test
whether the NLO BFKL formula (\ref{al1}) admits a phenomenological 
verification, 
i.e. whether $\g_c \approx \gamma^*.$ In other words we have to check 
the consistency condition expressed as 
\begin{eqnarray}
\chi_{eff} (\gamma^* (\om,Q^2),\al_{RG}(Q^2)) = 
\frac{\om}{\al_{RG}(Q^2)}\ .
\label{chilof}
\end{eqnarray}
 We considered  (\ref{chilof})  following the two 
resummation
schemes defined in section II. For scheme $S3$ 
$\chi^{eff}$ as a function of $\om$ is given in 
Fig.\ref{chinlos3}. Note that the 
different parametrisations agree except at high values of $Q^2,$ which 
are anyway 
out of the scope of our BFKL study.

Interestingly, while the scheme $S4$ (not shown in Fig.\ref{chinlos3}) gives 
the 
same 
curves as for $S3,$  the same test could not be safely performed for the 
$CCS$ 
scheme. 
The reason is that a spurious pole appears in the quark sector at $\om=1,$ 
when 
$\gamma^* \ne 0$  due to the incomplete momentum sum rule  in the 
quark 
sector, 
which gives an effective anomalous dimension slightly below the  value $0$ at 
$\om=1$. 
This is a technical difficulty which requires  a better treatment 
of 
the 
quark and gluon sectors simultaneously, which is still a challenge in the NLO 
BFKL 
approaches \cite{salambis}.
 
\begin{figure}
\begin{center}
\epsfig{figure=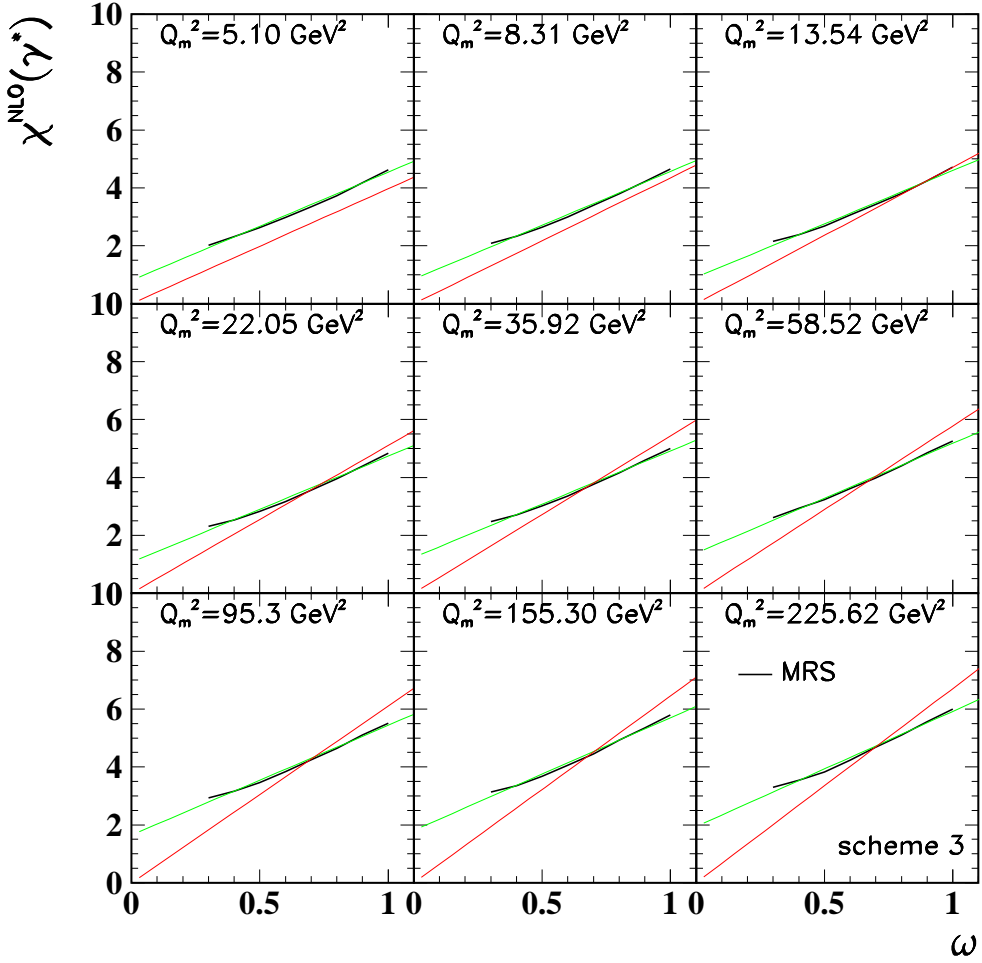,height=7.in}
\end{center}
\caption{{\it Phenomenological analysis of the  effective kernel in 
Mellin space.} 
Continuous lines: \ $\chi_{eff} $. Dashed lines: linear 
approximation for  
as a function of $\om.$ for the different bins of $Q^2$ and 
for scheme $S3.$ Continuous straight lines: consistency condution (\ref{chilof}), 
see text}
\label{relations3}
\end{figure}

We notice in Fig.\ref{chinlos3} that the linear property of relation 
(\ref{chilof}), namely for $\chi_{eff} (\gamma^* 
(\om,Q^2),\al_{RG}(Q^2))$ as a 
function of $\om$ 
(\ref{chilof})
is well verified. We indeed  can describe
the GRV and MRS parametrisations using a linear fit with a good 
precision. However 
the predicted zero at the origin   $\om =0$ is not obtained, even if the 
value at 
the origin 
remains small. The fit does not go through the origin and we would need to 
add 
a 
constant
term to the linear fit formula. In order to quantify the observed 
discrepancy, we 
reformulated the phenomenologically obtained curves by the following 
formula:
\begin{eqnarray}
\chi^{NLO} (\gamma^* (\om,Q^2), {\alpha}_{in} ) = 
{\om}/{{\alpha}_{out}}\ ,
\label{inout}
\end{eqnarray}
where ${\alpha}_{in}\equiv {\alpha}_{RG}$ is the theoretical input 
(\ref{al}) 
while ${\alpha}_{out}$ is the phenomenologically determined slope of the 
linear fit
displayed in Fig.\ref{relations3}. The validity of (\ref{chilof}) would 
obviously 
require ${\alpha}_{in}\equiv{\alpha}_{out}.$ 
In order to make the comparison, we have 
drawn the straight lines going from the origin, corresponding to the consistency 
condition  (\ref{chilof}).

\begin{figure}
\begin{center}
\epsfig{figure=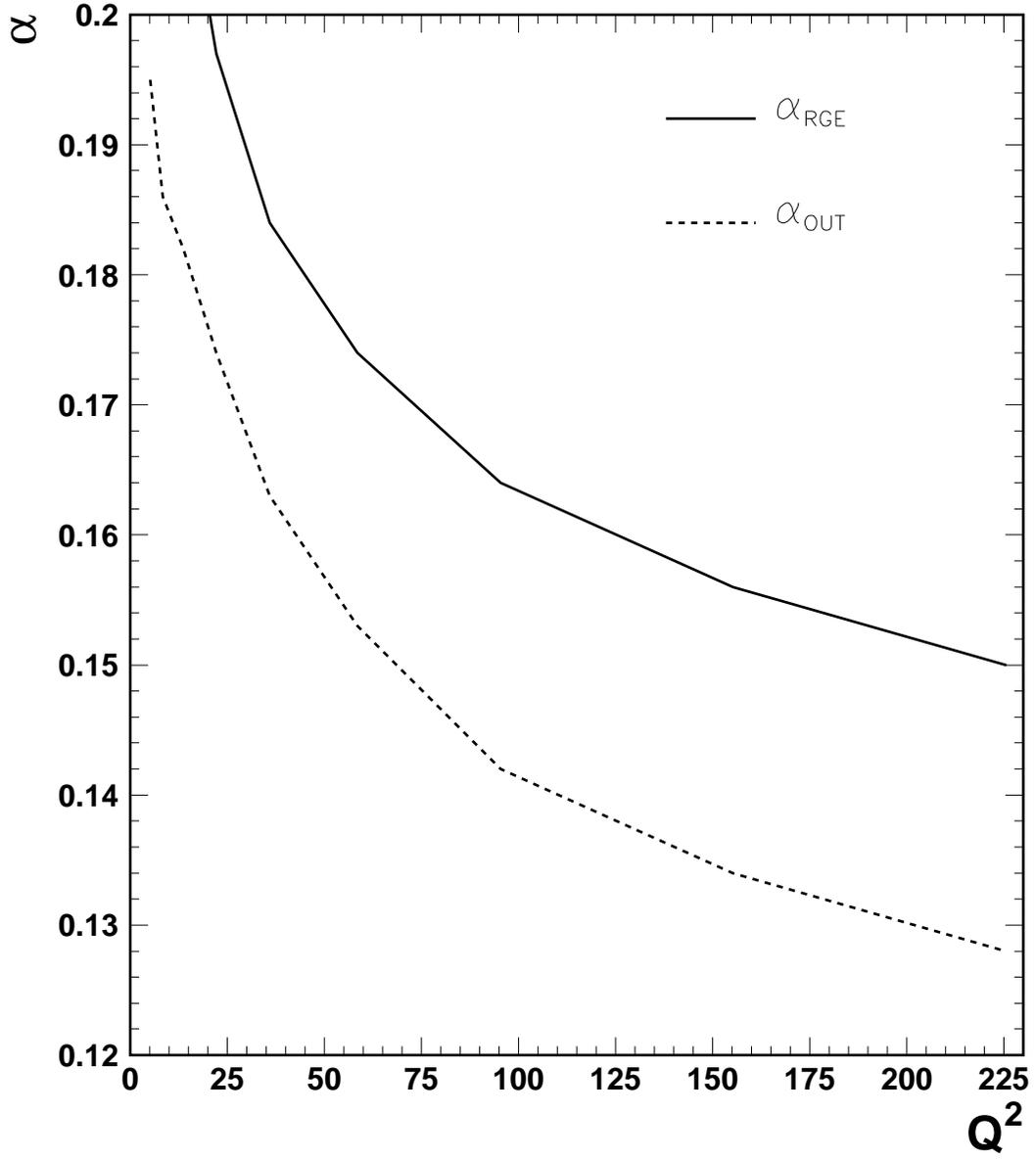,height=7.in}
\end{center}
\caption{$\alpha$ as a function of $Q^2$: Obtained by RGE (upper curve)
or as an output of the fit (lower curve) (see text).}
\label{alpha}
\end{figure}

We show in Fig.\ref{alpha}, the values of
$\alpha_{in}$ using the RGE equation (upper curve), and the values of
$\alpha_{out}$ (lower curve), which are always smaller than the values 
of
$\alpha_{RG}$.   $\alpha_{out}$ is
found to be closer to $\alpha_{in}$ at low $Q^2$ but more different from 
this value
at higher values of $Q^2$. 

\begin{table} [t]
\begin{center}
\begin{tabular}{|c||c|c|c|c|c|c|} \hline
BFKL fit & $\alpha$ & $Q_0^2$ &  ${\cal N}$ 
&
$\chi^2$ ( (/dof)) \\
\hline\hline
 LO  & 0.092 $\pm$ 0.010 
 & 0.401 $\pm$ 0.012 
 & .103 $\pm$ .002 
 
 & 1.33 (70)\\

 LO' (${\alpha}_{RG}$) & ------ 
  & .055$\pm$ 0.23 
  & 1.055  $\pm$ .032 
  & 8.93 (71)\\

 NLO (S3) & ------ 
  &  3.39 $\pm$ 0.23 
  & .101  $\pm$ .003 
  & 3.13 (71)\\
  
   NLO (CCS) & ------
  & 4.27 $\pm$ 0.30
  & .091  $\pm$ .007
  & 8.62 (71)\\

\hline
\end{tabular}
\caption{{\it Results of the BFKL fits to the H1 data.}
LO BFKL ($\bal = cst.$):  (1rst line); LO BFKL (${\alpha}_{RG}(Q^2)$):  (2nd line);NLO 
BFKL $S3$ scheme (3rd line); NLO BFKL $CCS$ scheme 
(4th 
line).
 }
 \end{center}
\label{table}
\end{table}
\section{Conclusion}
\label{5}
Summarizing the results of our paper, we have confronted the predictions of 
BFKL 
kernels 
at the level of leading and  
next-leading  logarithms (with resummation) with structure 
function data,  using two 
different proposed types  of resummation. Our method can be extended to other 
resummation proposals.

In a first stage we have proposed to use the ``effective kernel'' 
approximation of the NLO-BFKL kernels which, associated with the usual  
saddle-point approximation at high rapidity and large enough  $Q^2,$ allows 
one to obtain a simple two-parameter formula for the structure function 
$F_2.$ The comparison with the similar 3-parameter formula commonly used at 
LO level shows a deterioration of the fits when using two of the known 
resummed NLO schemes and a sensitivity to the different types   of 
resummation.

In order to look for a more model-independent discussion of the
discrepancy between precise data and the formulation in $x_{Bj}$-space, we 
perform an analysis of the kernel properties in Mellin-space. For this sake 
we find an interval in the energy-conjugate variable $\om$ where the 
different Mellin transformed analyses from data give a definite answer within 
reasonable error bars. In  Mellin-space, we find that small but sizeable 
effects give phenomenological deviations from the expected theoretical 
properties of the NLO kernels.

One possibility is that the saddle-point approximation we introduced is not 
valid. 
This 
simplicity assumption is phenomenologically motivated by its validity already 
at 
LO level. 
Unknown aspects of the 
prefactors, in particular the non-perturbative ones, could play a role in these 
deviations.

 One way out  is  to 
look 
for   higher order 
effects which could serve as a guide  to improved resummation procedures of 
NLO BFKL kernels. hence, it deserves to  investigate the phenomenological virtues of 
other proposed schemes 
and/or trying to make the suitable modifications to the known ones. Also, more 
precision in the 
discussion of NLO-BFKL predictions 
will soon be available with the completion of the perturbative NLO impact 
factors. 

A possibility is to incorporate subasymptotic effects which go beyond  the saddle-point 
approximation and may be computed from the NLO formulae 
\cite{altar,instabilities,salamalec}.
We expect 
our methods to be valid when incorporating these effects in the 
phenomenological analysis of $F_2$ data and thus further studies in the direction  
proposed in the present paper seem welcome.

\begin{acknowledgments}
R.P. wants to thank Gavin Salam for fruitful discussions and Jerome Salomez, 
Dionysis Triantafyllopoulos for the use of   Ref.\cite{trianta} and 
Ref.\cite{salomez} for the formulation  of the BFKL kernels.
Corrections of 
the 
manuscript by Rikard 
Enberg were welcome.
 \end{acknowledgments}

\appendix
\section*{Appendix}
\setcounter{section}{1}
\setcounter{equation}{0}
\numberwithin{equation}{section}

\subsection{NLO-BFKL Schemes $S3$ and $S4$ of Ref.\cite{salam}}
 
Let us  define the function
$$
4\chi_1(\gamma)=-\left[\frac{2\beta_0}{C_A}\chi_0(\gamma)^2-K\chi_0(\gamma)-
6\zeta(3)+\frac{\pi^2 
\cos(\pi\gamma)}{\sin^2(\pi\gamma)(1-2\gamma)}\left(3+\left(1+\frac{n_f}
{C_A^3}\right)\frac{2+3\gamma(1-\gamma)}{(3-2\gamma)(1+2\gamma)}\right)
-\right.
$$
\begin{equation}
\left. 
-\psi''(\gamma)-\psi''(1-\gamma)-\frac{\pi^3}{\sin(\pi\gamma)}+4\phi(\gamma)
\right]
\end{equation}
where 
\begin{equation}
\beta_0 =\frac{11C_A}{12}-\frac{2n_f}{12}\ ;
K=\frac{67}{9}-\frac{\pi^2}{3}-\frac{10n_f}{9C_A}
\end{equation}
 $C_A=3$ and  $n_f=3, $ 
\begin{equation}
\phi(\gamma)\equiv\sum_{n=0}^{\infty} (-1)^n 
\left[\frac{\psi(n+1+\gamma)-\psi(1)}{(n+\gamma)^2}+\frac{\psi(n+2-\gamma)
-\psi(1)}{(n+1-\gamma)^2} \right]
\end{equation}
and
\begin{equation}
\chi_0(\gamma)=2\psi(1)-\psi(\gamma)-\psi(1-\gamma)\ .
\end{equation}
 We 
consider the asymptotic expansion of  $\chi_1(\gamma)$ in $0:$
$$
\chi_1(\gamma)=-\frac{1}{4}
\left(\left(\frac{2\beta_0}{C_A}+3+\frac{2}{3}
\left(1+\frac{n_f}{C_A^3}\right)\right)\frac{1}{\gamma^2}+
\frac{2}{\gamma^3}
+\left(-K+\frac{67+13\frac{n_f}{C_A^3}}{9}-\pi^2+4\psi'(1)\right)
\frac{1}
{\gamma}-6\zeta(3)+\right.
$$
$$
\left.+\ 
\frac{395+71\frac{n_f}{C_A^3}}{27}-\frac{\pi^2}{2}-2\psi''(1)+2\psi''
(1)\right)+O(\gamma)=-\frac{1}{2\gamma^3}-\left(\frac{\beta_0}{2C_A}+
\frac{3}{4}+\frac{1}{6}\left(1+\frac{n_f}{C_A^3}\right)\right)\frac{1}
{\gamma^2}+
$$
\begin{equation}
+\left(\frac{K}{4}-\frac{67+13\frac{n_f}{C_A^3}}{36}+\frac{\pi^2}
{4}-\psi'(1)\right)\frac{1}{\gamma}\frac{3}{2}\zeta(3)-
\frac{395+71\frac{n_f}{C_A^3}}{108}+\frac{\pi^2}{8}+
O(\gamma)\ ,
\end{equation}
and  denote 
 $d_{1,k}$ the coefficient of $\frac{1}{\gamma^k}$ in this expansion. In 
particular we get
\begin{equation}d_{1,1} 
=-\frac{67}{36}-\frac{13n_f}{36C_A^3}+\frac{\pi^2}{12}+\frac{K}{4} \ ; 
d_{1,2} =-\frac{\beta_0}{2C_A}-\frac{11}{12}-\frac{n_f}{6C_A^3} \ ;
d_{1,3} =-\frac{1}{2}\ .
\end{equation}\\

We write
\be
\chi^{(0)}(\gamma)=\chi^{(0)}(\gamma,\omega=\bar{\al}\chi^{(0)})=
(1-\bar{\al}A)
\ \left(2\psi(1)-\psi(\gamma+\frac{1}{2}\omega+\bar{\al} 
B)-\psi(1-\gamma+\frac{1}{2}\omega+\bar{\al} B )\right)
\label{chi}\ee
where  $ A,B$  are constants. The  coefficient of the term 
in $\bar{\al}$  in the expansion  of  $\chi^{(0)}$ is
\begin{equation}
\chi_1^{(0)}(\gamma)=-\left(B+1/2 \chi_0(\gamma)\right)(\psi'(\gamma)
+\psi'(1-\gamma))-A\chi_0(\gamma)\ ,
\end{equation}
whose asymptotic expansion in $0$ is given by
$$
\chi_1^{(0)}=-(B+\frac{1}{2\gamma}+O(\gamma^2))(\frac{1}{\gamma^2}+2\psi'(1-
\gamma)+O(\gamma))-A(\frac{1}{\gamma}+O(\gamma^2))-\frac{1}{2\gamma^3}-\frac
{B}{\gamma^2}-\frac{A+\frac{\pi^2}{6}}{\gamma}+O(1)\,.
$$
Denoting $d^{(0)}_{1,k}$ the coefficient of $\frac{1}{\gamma^k}$ , one 
obtains
\begin{equation}d^{(0)}_{1,1} =-A-\frac{\pi^2}{6}\ ;
d^{(0)}_{1,2} =-B \ ; 
d^{(0)}_{1,3} =-\frac{1}{2}\ .
\end{equation}

Hence,
\begin{equation}
\chi_1(\gamma)-\chi_1^{(0)}(\gamma)= 
(d_{1,2}-d_{1,2}^{(0)})\frac{1}{\gamma^2}+(d_{1,1}-d_{1,1}^{(0)})\frac{1}
{\gamma}+O(1)\,.
\end{equation}
Choosing  A and B for eliminating spurious singularities in  $0:$
\begin{equation}
A  =-d_{1,1}-\frac{\pi^2}{6} \ ; B  =-d_{1,2}\ .
\end{equation}
we define 
\begin{equation}
\chi^{(1)'}(\gamma) 
=\chi^{(0)}(\gamma,\omega=\bar{\al}\chi^{(1)})=\chi^{(0)}(\gamma,\omega)+
\bar{\al}(\chi_1(\gamma)-\chi_1^{(0)}(\gamma))\,.
\end{equation}
The final NLO-BFKL kernel of $S3$ is finally given by 
performing the shift
\begin{equation}
\chi^{(1)}(\gamma)=\chi^{(1)'}(\gamma-\frac{1}{2}\omega)\,.
\label{shift}
\end{equation}\\

For the scheme $S4$ of Ref.\cite{salam}, instead of (\ref{chi}), one starts 
with 

\be
\chi^{(0)}(\gamma)=\chi^{(0)}(\gamma,\omega=\bar{\al}\chi^{(0)})
=\chi_0(\gamma)
-\frac{1}{\gamma}-\frac{1}{1-\gamma}
+(1-\bar{\al} A)\left[ 
\frac{1}{\gamma+\frac{1}{2}\omega+\bar{\al}B}+\frac{1}{1-\gamma+\frac{1}{2}
\omega+
\bar{\al}B}\right]\ .
\end{equation}

One has 
\ba
\chi_1^{(0)}(\gamma)&=&-A\left(\frac{1}{\gamma}+\frac{1}{1-\gamma}\right)
-\left(\frac{1}{2}\chi_0(\gamma)+B\right)\left(\frac{1}{\gamma^2}
+\frac{1}{(1-\gamma)^2}\right)= \no
&=& 
-A\left(\frac{1}{\gamma}+\frac{1}{1-\gamma}\right)-\left(\frac{1}{2\gamma}
+B\right)\left(\frac{1}{\gamma^2}
+\frac{1}{(1-\gamma)^2}\right)+O(1)
\ea
near 0. The singularity coefficients are now
\begin{equation}
d_{1,1}^{(0)}=-A-\frac{1}{2} \ ;\ d_{1,2}^{(0)}=-B \ ;\ 
d_{1,3}^{(0)}=-\frac{1}{2} 
\ .
\end{equation}
Thus, to cancel the divergences of $\chi_1(\gamma)-\chi_1^{(0)}(\gamma)$, one 
is 
led to choose 
\begin{equation}
A=-d_{1,1}-\frac{1}{2} \ ;\ B=-d_{1,2} \ .
\end{equation}
$\chi^{(1)}(\gamma)$ is calculated as for $S3.$\\

\subsection{NLO-BFKL Scheme $CCS$ \cite{autres} as defined in 
\cite{trianta}}

Starting   from the NLO-BFKL term $\chi_1(\gamma)$ as in [A.1],
one  first  introduces  the
DGLAP splitting functions and their Mellin transforms, which are
\begin{equation}\label{A3}
    P_{gg}(z)=\frac{z}{(1+z)_+}
    +\frac{1-z}{z}+z(1-z)
    +\f{\beta_0}{2C_A}\,\delta(1-z) \ ;\ 
    P_{qg}=\frac{N_f}{2N_c}[z^2+(1-z)^2]\ .
\end{equation}

Giving
\be
    A_{gg}(\omega)=
    \int_0^1 dz\, z^{\omega} P_{gg}(z)
    \; -\frac{1}{\omega}=\f{\beta_0}{2C_A}-\frac{1}{1+\omega}+
    \frac{1}{2+\omega}-
    \frac{1}{3+\omega}
    -[\psi(2+\omega)-\psi(1)]\ ,
\ee

\be
    A_{qg}(\omega) =
    \int_0^1 dz\, z^{\omega} P_{qg}(z)
    =\frac{N_f}{2N_c}
    \bigg(\frac{1}{1+\omega}-
    \frac{2}{2+\omega}+
    \frac{2}{3+\omega}
    \bigg)\ ,
\ee

\noindent one defines 

\begin{equation}\label{A7}
    A_{T}(\omega)= A_{gg}(\omega)
    +\frac{C_F}{N_c} A_{qg}(\omega),
\end{equation}

\noindent with $C_F=(N_c^2-1)/2N_c$. One can write the pole
structure of $\chi_1(\gamma)$ as

\be
    \chi_1(\gamma)=
    -\frac{1}{2 \gamma^{3}}
    -\frac{1}{2 (1- \gamma)^{3}}
    +\frac{A_{T}(0)}{\gamma^{2}}
    +\frac{A_{T}(0)-\f{\beta_0}{2C_A}}{(1-\gamma)^{2}}-F 
\bigg(\frac{1}{\gamma}+\frac{1}{1-\gamma}\bigg)
    +{\rm finite},
\ee

\noindent where $F$, which vanishes when $N_f=0$ causing the 
simple poles to go away, is

\begin{equation}\label{A9}
    F=\frac{N_f}{6N_c}
    \bigg(\frac{5}{3}+\frac{13}{6N_c^2}\bigg)\ .
\end{equation}
Following the same extraction of spurious poles as previously, we are led to 
define
\be
    \tilde{\chi}_1(\gamma)=
    \chi_1(\gamma)
    -\frac{1}{2}\chi_0(\gamma)\chi'_0(\gamma)
    +\frac{\chi_0(\gamma)}{(1-\gamma)^2}-\frac{1}{\gamma}
    +F \bigg(\frac{1}{\gamma}+\frac{1}{1-\gamma} \bigg)
    -\frac{A_T(0)}{\gamma^2}
    -\frac{A_T(0)-\f{\beta_0}{2C_A}}{(1-\gamma)^2}\ ,
\ee

\noindent a function with no poles at all at $\gamma=0$ and
$\gamma=1$.
Next, one  defines
\be
    \chi_1(\gamma,\omega)=
    \tilde{\chi}_1(\gamma)
    +\frac{1}{\gamma}
    -F \bigg(\frac{1}{\gamma}+\frac{1}
    {1-\gamma+\omega} \bigg)+\frac{A_T(\omega)}{\gamma^2}
    +\frac{A_T(\omega)-\f{\beta_0}{2C_A}}{(1-\gamma+\omega)^2}\ .
\ee

\noindent Finally the full kernel is given by
\begin{equation}\label{16}
    \chi(\gamma,\omega)=\chi_0(\gamma,\omega)+\omega\,
    \frac{\chi_1(\gamma,\omega)}{\chi_0(\gamma,\omega)}\ ,
\end{equation}
which has to be shifted as in (\ref{shift}).

\end{document}